\documentclass[letter,bibyear]{aa} 

\usepackage{epsfig}
\usepackage{amsmath}
\usepackage{subfigure}
\usepackage{natbib}
\usepackage{multirow}
\usepackage{color}
\usepackage{longtable}
\usepackage{graphicx}
\usepackage{epstopdf}
\usepackage{booktabs}
\usepackage{txfonts}

\newcommand{\cpl}{Chem. Phys. Lett.}

\newcommand{\jms}{J.~Mol.~Spectrosc.}   
\newcommand{\jmst}{J.~Mol.~Struct.}   

\newcommand{\kms}{km s$^{-1}$}

\newcommand{\once}{10$^{11}$\,cm$^{-2}$}
\newcommand{\doce}{10$^{12}$\,cm$^{-2}$}
\newcommand{\trece}{10$^{13}$\,cm$^{-2}$}
\begin{document}

\title{Pure hydrocarbon cycles in TMC-1: Discovery of ethynyl cyclopropenylidene, cyclopentadiene and indene.
\thanks{Based on observations carried out
with the Yebes 40m telescope (projects 19A003,
20A014, 20D023, and 21A011) and the Institut de Radioastronomie Millim\'etrique (IRAM) 30m telescope. The 40m
radiotelescope at Yebes Observatory is operated by the Spanish Geographic 
Institute
(IGN, Ministerio de Transportes, Movilidad y Agenda Urbana). IRAM is supported by INSU/CNRS
(France), MPG (Germany) and IGN (Spain).}}

\author{
J.~Cernicharo\inst{1},
M.~Ag\'undez\inst{1},
C.~Cabezas\inst{1},
B.~Tercero\inst{2,3},
N.~Marcelino\inst{1},
J.~R.~Pardo\inst{1},
P.~de~Vicente\inst{2}
}

\institute{Grupo de Astrof\'isica Molecular, Instituto de F\'isica Fundamental (IFF-CSIC),
C/ Serrano 121, 28006 Madrid, Spain\\ \email jose.cernicharo@csic.es
\and Centro de Desarrollos Tecnol\'ogicos, Observatorio de Yebes (IGN), 19141 Yebes, Guadalajara, Spain
\and Observatorio Astron\'omico Nacional (OAN, IGN), Madrid, Spain
}

\date{Received; accepted}

\abstract{We report the detection for the first time in space of three new pure hydrocarbon cycles
in TMC-1: $c$-C$_3$HCCH (ethynyl cyclopropenylidene), $c$-C$_5$H$_6$ (cyclopentadiene) and $c$-C$_9$H$_8$ 
(indene). We derive a column density of 3.1\,$\times$10$^{11}$ cm$^{-2}$ for the former cycle 
and similar values, in the range (1-2)\,$\times$10$^{13}$ cm$^{-2}$, for the two latter molecules. This 
means that cyclopentadiene and indene, in spite of their large size, are exceptionally abundant, only a 
factor of five less abundant than the ubiquitous cyclic hydrocarbon $c$-C$_3$H$_2$. The high abundance 
found for these two hydrocarbon cycles, together with the high abundance previously found for the 
propargyl radical (CH$_2$CCH) 
and other hydrocarbons like vinyl and allenyl acetylene \citep{Agundez2021,Cernicharo2021a,Cernicharo2021b}, 
start to allow us to quantify the abundant content of hydrocarbon rings in cold dark clouds and to identify 
the intermediate species that are probably behind the in situ bottom-up synthesis 
of aromatic cycles in these environments. 
While $c$-C$_3$HCCH is most likely formed through the reaction between the radical 
CCH and $c$-C$_3$H$_2$, the high observed abundances of cyclopentadiene and indene are difficult 
to explain through currently proposed chemical mechanisms. Further studies are needed to identify 
how are five- and six-membered rings formed under the cold conditions of clouds like TMC-1.}


\titlerunning{Hydrocarbon cycles in TMC-1}
\authorrunning{Cernicharo et al.}

\maketitle

\section{Introduction}
Since the hypothesis that polycyclic aromatic hidrocarbons (PAHs) are the 
carriers of 
the unidentified infrared bands \citep{Leger1984,Allamandola1985}, 
many efforts have been devoted to understand the 
chemical processes
leading to the formation of these molecular species (see, e.g., \citealt{Joblin2018}). 
Circumstellar envelopes around carbon-rich Asymptotic Giant Branch (AGB) stars have been suggested 
as the factories of PAHs \citep{Cherchneff1992}. 
The detection of benzene in the carbon-rich protoplanetary nebula CRL\,618 \citep{Cernicharo2001}
suggests a bottom-up approach in which the small hydrocarbons formed during the AGB phase, such 
as C$_2$H$_2$ and C$_2$H$_4$, interact with the ultraviolet (UV) radiation produced by the star 
in its evolution to the white
dwarf phase \citep{Woods2002,Cernicharo2004}. Other hypotheses involve the proccesing of dust 
grains around evolved stars, either through UV photons \citep{Pilleri2015} or by chemical 
processes \citep{Martinez2020}.
Hence, it has been surprising to see that cyanide derivatives of PAHs have been found in a 
cold prestellar core such as TMC-1, which is well protected against UV radiation 
\citep{McGuire2018,McGuire2021}. It is unlikely that these PAH-cyanides arise from a reservoir of
PAHs existing since the early stages of the cloud, since these relatively 
small PAHs would not have survived the diffuse cloud stage. 
Although \citet{McGuire2021} propose
a reasonable chemical network starting with the phenyl radical that could 
explain the observed abundances of
cyanonaphthalene, the chemical routes leading to benzene itself are still 
unclear. An in situ formation mechanism for benzene must involve abundant 
hydrocarbons containing from 2 to 4 carbon atoms. Moreover, some of these
species have to permit an easy cyclization in 2-3 steps to have an efficient yield of benzene or phenyl radical.
The propargyl radical, CH$_2$CCH, has been found recently in 
TMC-1 by \citet{Agundez2021} with an abundance close to 10$^{-8}$ relative to H$_2$. In addition, complex hydrocarbons
such as vinyl and allenyl acetylene have been also observed with very large abundances \citep{Cernicharo2021a,
Cernicharo2021b}. These hydrocarbons may hold the key 
to form the first aromatic ring, from which larger PAHs can grow.

\begin{figure*}[]
\centering
\includegraphics[scale=0.5,angle=0]{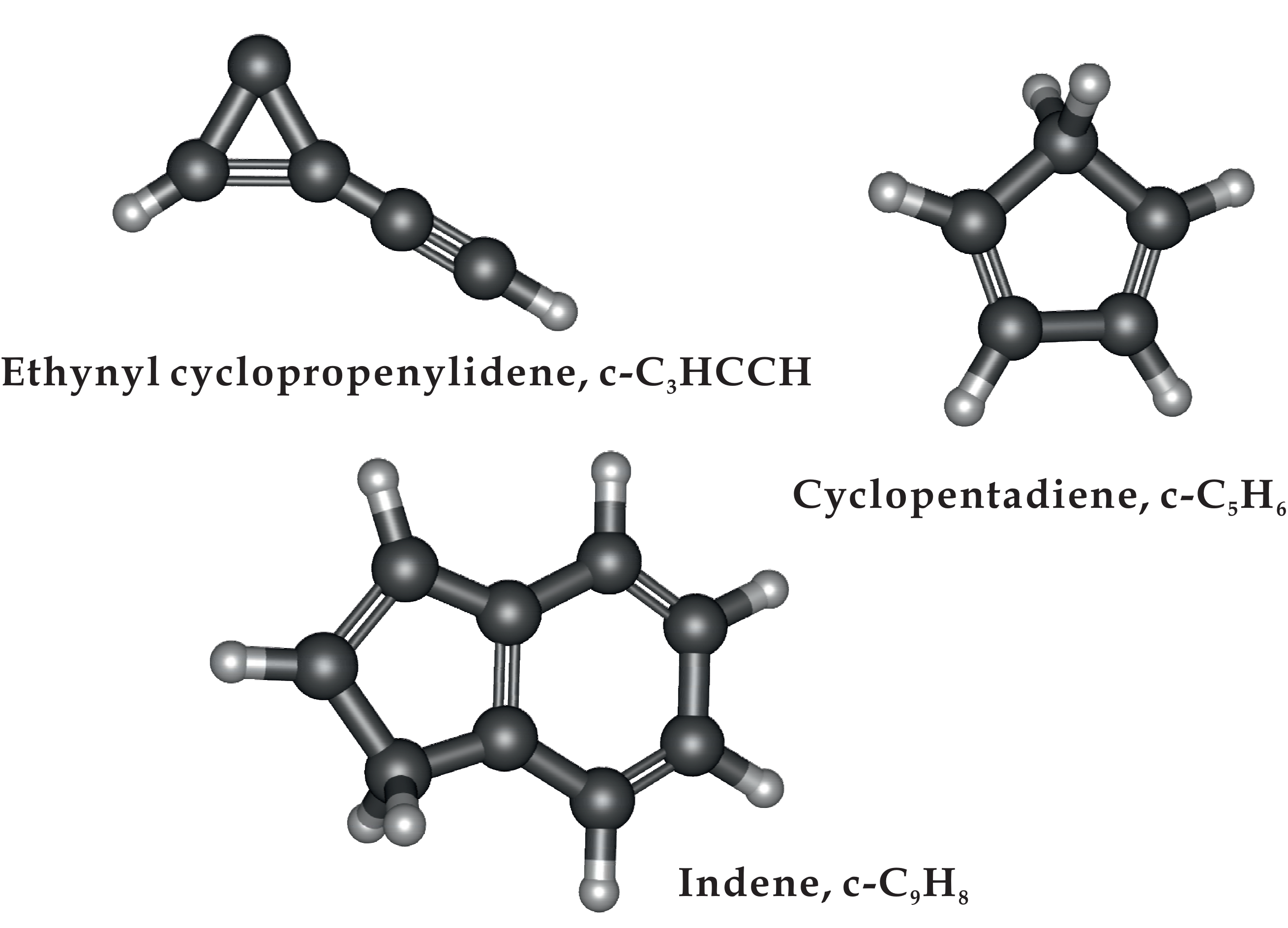}
\caption{Structures of the three species detected in this work.
}
\label{fig_carbon_cycles}
\end{figure*}

In this Letter we report the discovery of three pure hydrocarbon cycles 
(see Fig. \ref{fig_carbon_cycles}): $c$-C$_3$HCCH
(ethynyl cyclopropenylidene), $c$-C$_5$H$_6$
(cyclopentadiene), and $c$-C$_9$H$_8$ (indene)\footnote{We have learned in a virtual seminar
during the final phase of preparation of this work that a paper has been submitted (Burkhardt et al., 2021) claiming
the detection of this species in TMC-1. We do not know neither its content nor the quality of the
detection.}.

\begin{figure*}[]
\centering
\includegraphics[scale=0.65,angle=0]{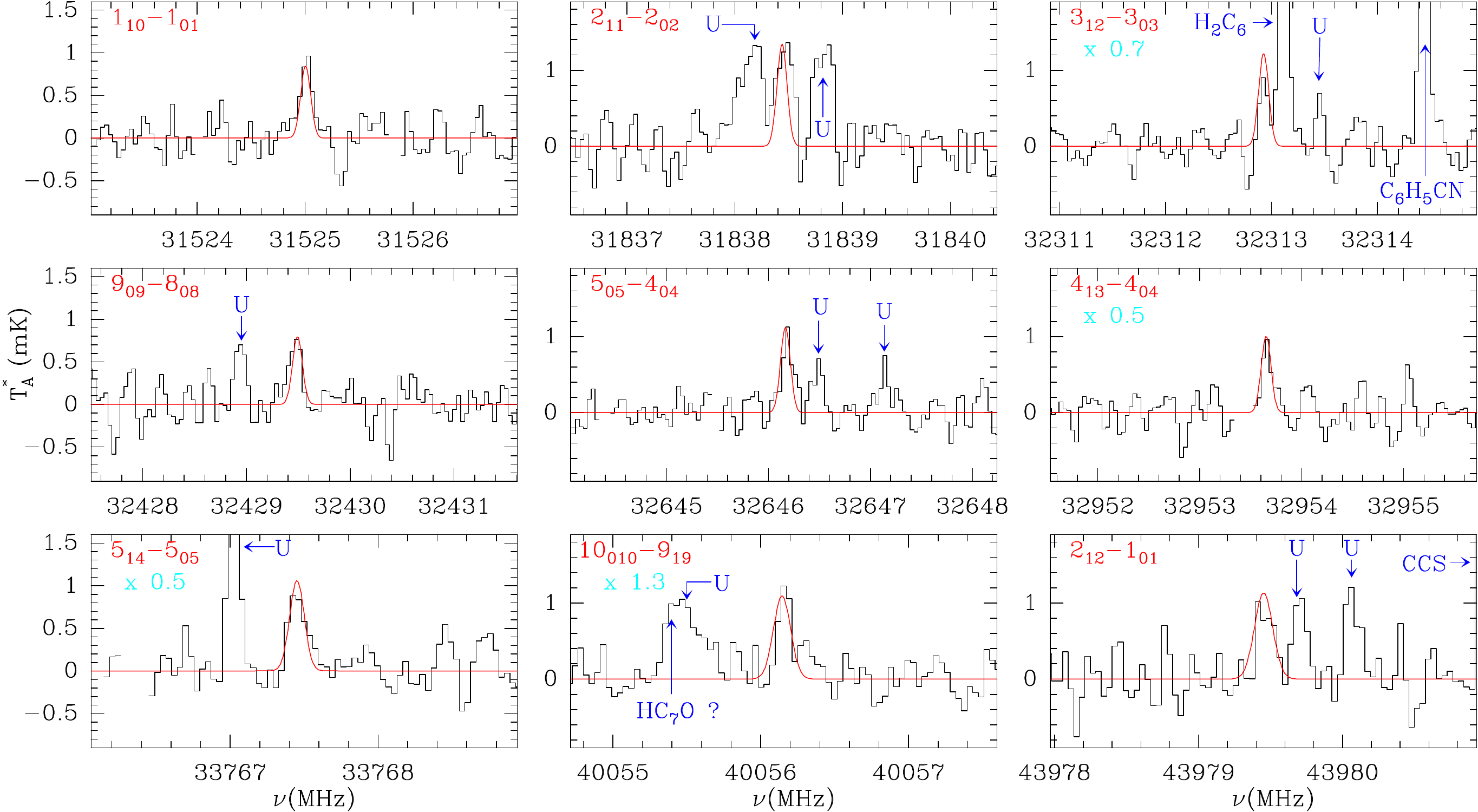}
\caption{Selected transitions of $c$-C$_3$HCCH in TMC-1.
The abscissa corresponds to the rest frequency of the lines assuming a
local standard of rest velocity of the source of 5.83 km s$^{-1}$. Frequencies and intensities for the observed lines
are given in Table \ref{obs_line_parameters}.
The ordinate is the antenna temperature, corrected for atmospheric and telescope losses, in milli Kelvin.
The quantum numbers for each transition are indicated
in the upper left corner of the corresponding panel.
The red lines show the computed synthetic spectrum for this species for T$_r$=10 K and
a column density of 3.1$\times$\once. 
Cyan labels indicated the multiplicative factor applied to the model to
match the intensity of the observed lines.
}
\label{fig_c3hcch}
\end{figure*}

\section{Observations}
\label{observations}
New receivers, built within the Nanocosmos project\footnote{\texttt{https://nanocosmos.iff.csic.es/}},
and installed at the Yebes 40m radiotelescope, were used
for the observations of TMC-1. 
The Q-band receiver consists of two HEMT cold amplifiers covering the 
31.0-50.3 GHz band with horizontal and vertical polarizations. Receiver temperatures vary from 22 K at 32 GHz 
to 42 K at 50 GHz. The backends are $2\times8\times2.5$ GHz fast Fourier transform spectrometers 
with a spectral resolution of 38.15 kHz 
providing the whole coverage of the Q-band in both polarisations. The main beam efficiency varies from 0.6 at 
32 GHz to 0.43 at 50 GHz. A detailed description of the system is given by \citet{Tercero2021}.

The line survey of TMC-1 ($\alpha_{J2000}=4^{\rm h} 41^{\rm  m} 41.9^{\rm s}$ and $\delta_{J2000}=+25^\circ 41' 27.0''$)
in the Q-band was performed in several sessions. Previous results on the detection of C$_3$N$^-$ and C$_5$N$^-$
\citep{Cernicharo2020b}, HC$_5$NH$^+$ \citep{Marcelino2020}, HC$_4$NC \citep{Cernicharo2020c}, and HC$_3$O$^+$
\citep{Cernicharo2020a} were based on two observing runs performed in November 2019 and February 2020. Two
different frequency coverages were used, 31.08-49.52 GHz and 31.98-50.42 GHz, in order to check that no
spurious spectral ghosts are produced in the down-conversion chain.
Additional data were taken in October, December 2020 and January-April 2021. 
The observing procedure was frequency-switching with a frequency throw of 10\,MHz for the two first runs and of 
8\,MHz for all the others.

All data were analyzed using the GILDAS
package\footnote{\texttt{http://www.iram.fr/IRAMFR/GILDAS}}.

\section{Results}
\label{results}
The sensitivity of our TMC-1 data is better than
previously published line surveys of this source at the same frequencies \citep{Kaifu2004} by a factor 10-20. 
In fact, it has been possible to detect many individual lines from molecules
that were reported previously only by stacking techniques \citep{Marcelino2021}. 
The recent discovery of some molecules
containing the ethynyl group (CCH), such as vinyl and allenyl acetylene \citep{Cernicharo2021a,Cernicharo2021b},
and of the propargyl radical \citep{Agundez2021},
prompted us to search for other chemically related hydrocarbons.
Line identification in this work was done using the catalogues 
MADEX \citep{Cernicharo2012}, CDMS \citep{Muller2005}, and JPL \citep{Pickett1998}.

\subsection{Ethynyl cyclopropenylidene, $c$-C$_3$HCCH}
\label{c-c3hcch}
Cyclopropenylidene ($c$-C$_3$H$_2$) is an abundant species in cold dark clouds. 
Several lines of the para species of this molecule have been detected in our line
survey. They are analyzed in Appendix~\ref{Ap_C3H2}. The reaction of this molecule
with CCH and CN could lead to the formation of $c$-C$_3$HCCH and c-C$_3$HCN.
Ethynyl cyclopropenylidene ($c$-C$_3$HCCH; see Fig. \ref{fig_carbon_cycles}) is one of the various C$_5$H$_2$ isomers 
(see Appendix \ref{Ap_isomers_C3HCCH})
studied 
by microwave spectroscopy in the laboratory \citep{Travers1997,McCarthy1997,Gottlieb1998}.
In addition to $c$-C$_3$HCCH, these isomers are $l$-H$_2$C$_5$, the bent HCC(CH)CC, and the
cycle $c$-H$_2$C$_3$CC. All these species are implemented in the MADEX code \citep{Cernicharo2012}. 
In the case of $c$-C$_3$HCCH, 
the dipole moment components are $\mu_a$=2.04\,D, and $\mu_b$=2.89\,D \citep{Travers1997}.

A search for $c$-C$_3$HCCH resulted in the detection of 13 rotational lines, which gives
a high degree of confidence in the detection of this species. Nevertheless,
the lines are weak, with intensities around 1 mK. Some of them are shown in Fig.~\ref{fig_c3hcch}.  
The derived line parameters (see Appendix~\ref{line_parameters}) are given in Table~\ref{obs_line_parameters}.
A few lines that could have been detected in our survey appear blended with lines from other 
species or are affected by negative features produced in the folding of frequency switching
procedure. These cases are indicated in Table~\ref{obs_line_parameters}.
An analysis of the observed intensities through a rotational diagram
provides a rotational temperature of 8\,$\pm$\,3\,K and a column density 
for $c$-C$_3$HCCH of 
(3.1\,$\pm$\,0.8)\,$\times$\,\once.
We assumed a linewidth of 1.0 \kms, which corresponds to the observed average value,
and a source of uniform brightness temperature with
a diameter of 80$''$ \citep{Fosse2001}.
We performed a model fitting directly
the observed line profiles as described  by \citet{Cernicharo2021c}, with the result that
the best match between the computed synthetic spectrum and the observations corresponds
to T$_r$\,=\,10\,K and a column density similar to that derived from the rotation diagram. 
Figure \ref{fig_c3hcch} shows, in red,
the computed synthetic spectrum. The correction factors applied to each line to match the observations
are indicated in the Figure. Taken into account the signal-to-noise ratio 
of the data,
these factors are within the expected uncertainties of the fit.

Adopting a column density of H$_2$ of 10$^{22}$ cm$^{-2}$ for TMC-1 \citep{Cernicharo1987}, the abundance of $c$-C$_3$HCCH relative
to H$_2$ is 3.1\,$\times$\,10$^{-11}$, which is significantly below those 
of $c$-C$_3$H$_2$ and $c$-C$_3$H
(5.9\,$\times$\,10$^{-9}$ and 6.2\,$\times$\,10$^{-10}$, respectively; see
Appendices~\ref{Ap_C3H2} and \ref{Ap_C3H}), ethynyl derivatives such as vinyl, methyl, and allenyl acetylene 
\citep{Cernicharo2021a,Cernicharo2021b,Cabezas2021}, 
and propylene \citep{Marcelino2007}. 
We derive the following abundance ratios: $c$-C$_3$H$_2$/$c$-C$_3$HCCH $\sim$\,190 (see Appendix~\ref{Ap_C3H2}), 
$c$-C$_3$H/$c$-C$_3$HCCH $\sim$\,20, and $c$-C$_3$H$_2$/$c$-C$_3$H $\sim$\,10. Upper 
limits to the column densities of the isomers of $c$-C$_3$HCCH
are analyzed in Appendix \ref{Ap_isomers_C3HCCH}. Improved rotational and 
distortion constants
for $c$-C$_3$HCCH are given in Appendix \ref{improved_rotational_constants}.

\begin{figure}[]
\centering
\includegraphics[scale=0.6,angle=0]{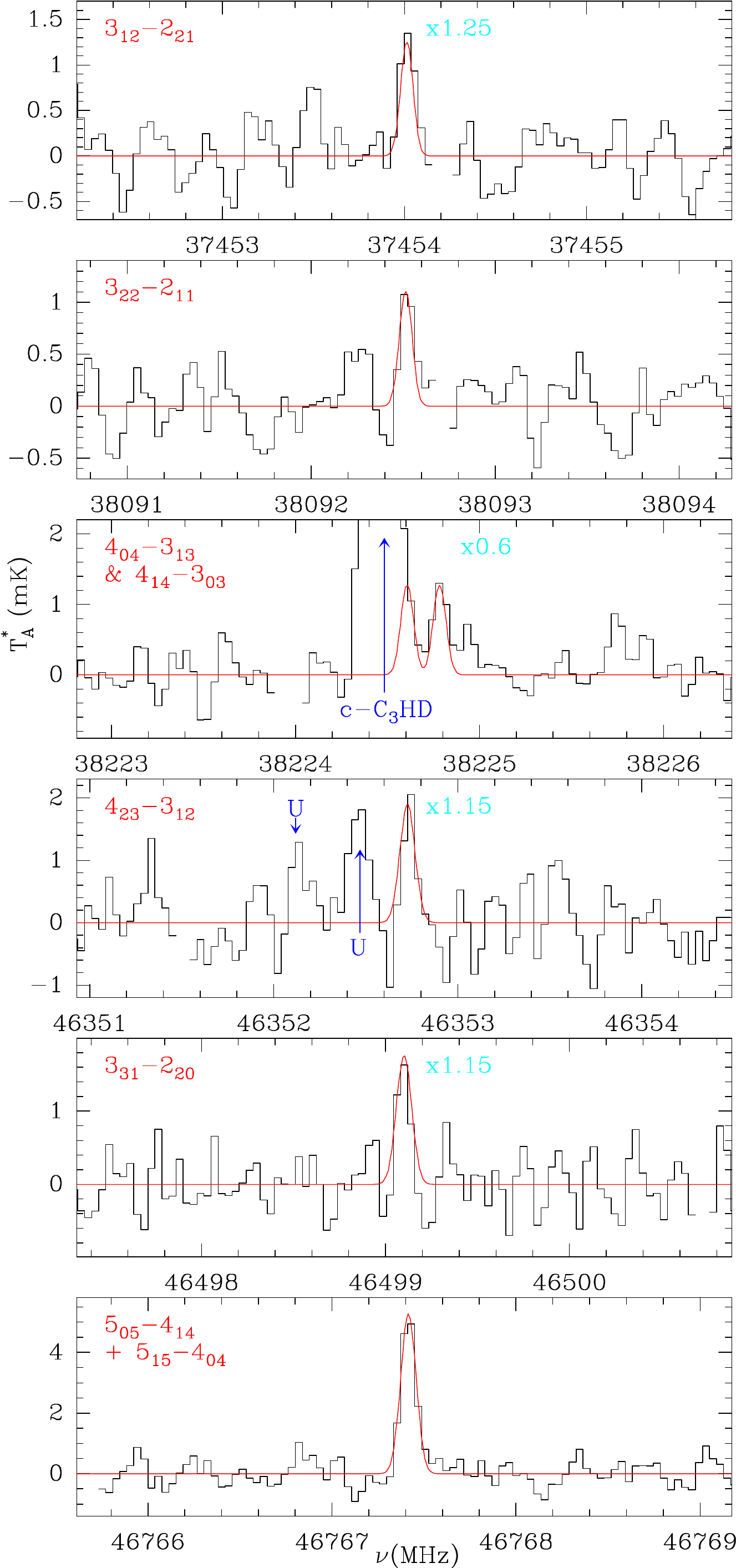}
\caption{Same as Fig. \ref{fig_c3hcch} but for the observed transitions of $c$-C$_5$H$_6$ towards TMC-1.
The red line shows the computed synthetic spectrum for cyclopentandiene assuming $T_r$\,=\,10\,K and
$N$($c$-C$_5$H$_6$)\,=\,1.2\,$\times$\,\trece. Cyan labels, when present, indicate the multiplicative factor applied to
the best fit model to match the observations.}
\label{fig_c-c5h6}
\end{figure}

\begin{figure*}[]
\centering
\includegraphics[scale=0.6,angle=0]{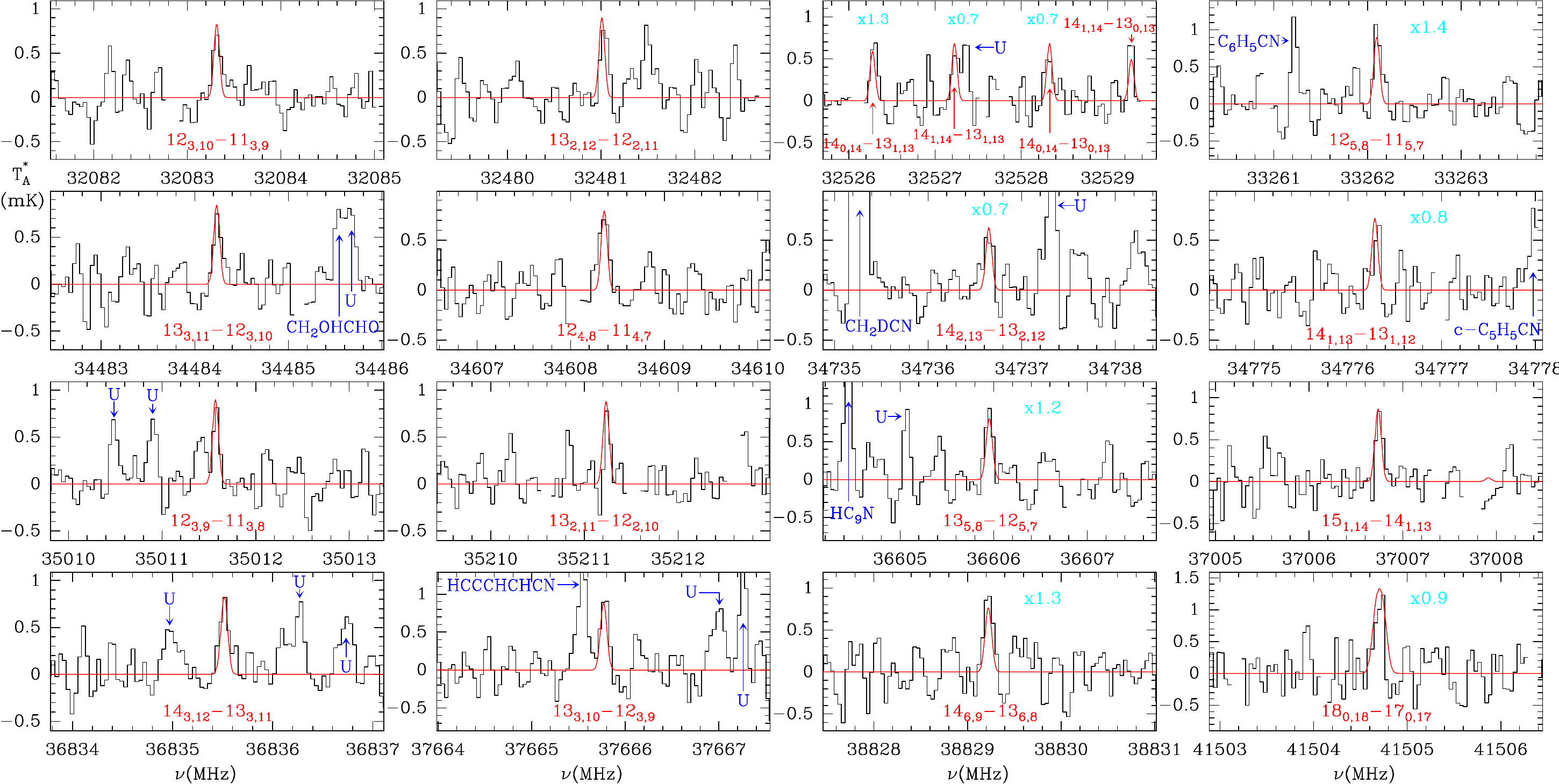}
\caption{Same as Fig. \ref{fig_c3hcch} but for the selected transitions of $c$-C$_9$H$_8$ observed towards TMC-1.
The red line shows the computed synthetic spectrum for indene assuming $T_r$\,=\,10\,K and
$N$($c$-C$_9$H$_8$)\,=\,1.6\,$\times$\,\trece. Cyan labels, when present, indicate the multiplicative factor applied to
the best fit model to match the observations.}
\label{fig_c-c9h8}
\end{figure*}

\subsection{Cyclopentadiene, $c$-C$_5$H$_6$}
The recent detection
by stacking techniques of two cyanide derivatives of cyclopentadiene in TMC-1 \citep{Lee2021} 
makes very likely that cyclopentadiene itself is also present in this source.
Cyclopentadiene (see Fig. \ref{fig_carbon_cycles}) has been observed in the laboratory up to 386.6 GHz by different
authors \citep{Laurie1956,Scharpen1965,Benson1970,Bogey1988}. The molecule has 
a low dipole moment of 0.416\,D along its $b$ axis \citep{Laurie1956}. 
Because of the C$_{2v}$ symmetry
of the molecule, two spin species, ortho and para, with $K_a+K_c$ odd and 
even, and statistical weights 9 and 7, respectively, have 
to be considered. The lowest ortho level is the 1$_{1,1}$ and it is 0.6 K above the lowest para
level (0$_{0,0}$). We searched for this species in our TMC-1 data and
we detected all the strong lines in the 31-50\,GHz frequency range. 
They are shown in Fig. \ref{fig_c-c5h6} and the line parameters are given 
in Table \ref{obs_line_parameters}.
Only one para line, the
$4_{1,3}-3_{2,2}$, which is affected
by a negative frequency-switching artifact, is missing. Another
para line, the $4_{0,4}-3_{1,3}$ is strongly blended with a line from $c$-C$_3$HD (see third panel from the top
in Fig. \ref{fig_c-c5h6}).
Taking into account the large geometrical section of the molecule and the 
low dipole moment, we could expect
to have its rotational levels thermalized at the kinetic temperature of the cloud, 10\,K. Assuming the
same source parameters than for $c$-C$_3$HCCH we derive a total column density for cyclopentadiene of 
(1.2\,$\pm$\,0.3)\,$\times$\,\trece, with identical contributions from the ortho and para species. 
The fractional abundance of 
 $c$-C$_5$H$_6$ relative to H$_2$ is thus 1.2\,$\times$\,10$^{-9}$. This abundance is identical 
 to that of CH$_2$CHCCH \citep{Cernicharo2021b} 
and H$_2$CCHCCH \citep{Cernicharo2021c}, and lower than that of the propargyl radical by a factor 
of $\sim$7 \citep{Agundez2021}. Compared with its cyano-derivatives \citep{Lee2021}, cyclopentadiene is
15 and 63 times more abundant that 1- and 2-cyanocyclopentadiene, respectively.

\subsection{Indene, $c$-C$_9$H$_8$}
Indene, $c$-C$_9$H$_8$, has a bicyclic structure with a six-membered ring 
fused to a five-membered ring (see Fig. \ref{fig_carbon_cycles}).
It has been studied in the microwave laboratory by \cite{Li1979} and \cite{Caminati1993}. 
The accuracy of these measurements is moderate, with typical uncertainties for the measured
frequencies of $\sim$100 kHz. Its
dipole moment is also relatively low, with $\mu_a$\,=\,0.50\,D and $\mu_b$\,=\,0.37\,D \citep{Caminati1993}.
Frequency predictions in the 31-50 GHz range have uncertainties between 10 and 80 kHz.
We searched for indene in our data and found a large number of lines above the 3$\sigma$ level. Nineteen 
of them
are shown in Fig. \ref{fig_c-c9h8}. Line parameters for all the 47 observed lines of indene detected in TMC-1
are given in Table \ref{obs_line_parameters}. Improved rotational and distortion constants
for indene are provided in Appendix \ref{improved_rotational_constants} and
Table \ref{rotational_constants_indene}.

In order to derive a column density for indene we have assumed
that the rotational temperature is 10\,K. Taking into account the large geometrical section of the
molecule and its low dipole moment, this assumption seems reasonable. The 
model fit procedure,
consisting in a fit to the observed line profiles, provides a column density for indene of
(1.6\,$\pm$\,0.3)\,$\times$\,\trece, which is very similar to that of cyclopentadiene, vinyl and allenyl
acetylene, and only a factor of five below that of $c$-C$_3$H$_2$. 
The fractional abundance of indene relative to H$_2$ is thus 1.6\,$\times$\,10$^{-9}$. 
The discovery of this large hydrocarbon through the old fashion line-by-line detection 
procedure is solid and robust. 
For this species, the observed averaged value of the linewidth is close to 1.0 \kms, 
such as for c-CH$_3$CCH. Most lines of polar species have linewidths of  0.6 \kms. In the case of indene, 
its low dipole moment and its easy to be thermalized conditions, could favour the emission from the less
dense regions of the cloud. This could suggest that this species prevails in the
molecular envelope of TMC-1. However, c-CH$_3$CCH has a larger dipole moment and requires
a high density to be close to thermalization as derived in section 3.1. Further observations
with higher spectral resolution are needed to discriminate between excitation and possible
contribution from different cloud components.

\section{Discussion} 
\label{discussion}

\subsection{$c$-C$_3$HCCH}

The most obvious route to $c$-C$_3$HCCH is the reaction between CCH 
and $c$-C$_3$H$_2$. To our knowledge, this reaction has not been studied either experimentally 
or theoretically, but it is likely that it occurs fast at low temperatures with the H loss 
channel being the major one, based on the reactivity of CCH with other unsaturated 
hydrocarbons \citep{Vakhtin2001}. In addition, the two reactants are quite abundant 
in TMC-1 (e.g., \citealt{Agundez2013}). If we implement this reaction with a rate 
coefficient of 10$^{-10}$ cm$^3$ s$^{-1}$ in a chemical model 
similar to that presented by \cite{Agundez2021} and we assume that $c$-C$_3$HCCH 
is mostly destroyed by reactions with C atoms and C$^+$ and H$^+$ 
ions, the peak abundance calculated for $c$-C$_3$HCCH is $\sim$\,5\,$\times$10$^{-11}$ 
relative to H$_2$, which is in very good agreement with the 
value observed in TMC-1. Although the subset of reactions involving $c$-C$_3$HCCH is 
clearly incomplete, this exercise shows that the reaction CCH 
+ $c$-C$_3$H$_2$ could be a plausible formation route for this cyclic hydrocarbon in 
TMC-1. Other reactions that could also yield $c$-C$_3$HCCH are $c$-C$_3$H + C$_2$H$_2$ 
and C$_2$ + CH$_2$CCH, although little is known 
on their chemical kinetics.

Similarly to CCH, the radical CN has been also found to react rapidly with unsaturated 
hydrocarbons at low temperatures (e.g., \citealt{Sims1993}). Therefore, the reaction between 
CN and $c$-C$_3$H$_2$ is a very likely source of $c$-C$_3$HCN, which is the cyanide derivative 
analogue of $c$-C$_3$HCCH. It is interesting to note that ethynyl derivatives are more 
abundant than cyanide derivatives in TMC-1, e.g., CH$_3$CCCCH/CH$_3$CCCN = 7.6 and 
CH$_2$CCHCCH/CH$_2$CCHCN = 4.4 \citep{Marcelino2021,Cernicharo2021b}, which is 
thought to reflect the CCH/CN abundance ratio of 10 observed in this source 
\citep{Pratap1997}. Therefore, we could expect $c$-C$_3$HCN to be a few times 
less abundant than $c$-C$_3$HCCH. We do not detect $c$-C$_3$HCN in our sensitive 
TMC-1 data, which means that it must have a column density below 
2.5\,$\times$\,10$^{11}$ cm$^{-2}$ (3$\sigma$ upper limit; see 
Appendix~\ref{Ap_isomers_C3HCCH}). Therefore, the abundance ratio 
$c$-C$_3$HCCH/$c$-C$_3$HCN in TMC-1 must be $>$ 1.2. It is therefore 
likely that a reduced noise level from deeper observations  
may lead to the detection of $c$-C$_3$HCN in TMC-1.

\subsection{Cyclopentadiene, $c$-C$_5$H$_6$}
Given the high abundance derived for cyclopentadiene 
in TMC-1, 1.2\,$\times$\,10$^{-9}$ relative to H$_2$, it is not straightforward 
to find an efficient formation route. Several reactions that could 
lead to $c$-C$_5$H$_6$ are not efficient under the cold conditions of TMC-1. 
For example, the reaction between CH$_2$CCH + C$_2$H$_4$ has an activation 
barrier \citep{Saeys2003}, while the reactions of C$_2$H$_3$ with either 
CH$_3$CCH or CH$_2$CCH$_2$ probably have also barriers, as occurs in other 
reactions of C$_2$H$_3$ with unsaturated hydrocarbons \citep{Miller2000,Ismail2007,Goldsmith2009}. 
The reaction between C$_3$H$_5$ and C$_2$H$_2$ yields cyclopentadiene, but it has a barrier 
\citep{Bouwman2015}. 
The reaction C$_2$H + CH$_2$CHCH$_3$ is fast at low temperatures and produces several 
C$_5$H$_6$ isomers, although the experimental data seems to be inconsistent with cyclopentadiene 
being one of them \citep{Bouwman2012}.

The reaction between CH and butadiene (CH$_2$CHCHCH$_2$) is calculated to produce 
$c$-C$_5$H$_6$ with no barrier \citep{McCarthy2021}. Adopting a rate coefficient of 
10$^{-10}$ cm$^3$ s$^{-1}$ for this reaction in our chemical model, based on \cite{Agundez2021}, 
and assuming that c-C5H6 is mostly removed through reactions with C atoms and C$^+$ 
and H$^+$ ions, we calculate a peak abundance of $\sim$\,3\,$\times$\,10$^{-13}$ 
relative to H$_2$ for $c$-C$_5$H$_6$. This is more than three orders of magnitude 
below the observed value.
In the chemical model, 
butadiene is essentially formed by the reaction CH + CH$_2$CHCH$_3$ 
\citep{Daugey2005,Loison2009,Ribeiro2016} with a peak abundance of $\sim$\,3\,$\times$10$^{-10}$ 
relative to H$_2$. Butadiene is an interesting potential precursor of $c$-C$_5$H$_6$, and also 
of benzene \citep{Jones2011}, but to play such a role it needs to have an abundance much higher 
than calculated by current gas-phase chemical models. Unfortunately, it is a non-polar molecule 
and thus it cannot be detected through radio techniques.

Given the high abundance derived for the radical CH$_2$CCH \citep{Agundez2021}, reactions between radicals, 
such as CH$_2$CCH + C$_2$H$_5$ and C$_2$H$_3$ + CH$_2$CHCH$_2$, are also potential sources of 
$c$-C$_5$H$_6$. 
Routes involving ions, such as condensation rections between cationic and neutral 
hydrocarbons \citep{Herbst1989}, may also lead to $c$-C$_5$H$_6$.
Theoretical studies on these reactions would help to evaluate their role in 
the synthesis of cyclopentadiene. Whatever the formation route to $c$-C$_5$H$_6$ is, 
this hydrocarbon 
cycle is the most obvious 
precursor of the two cyanide derivatives $c$-C$_5$H$_5$CN detected in TMC-1 \citep{McCarthy2021,Lee2021}.

\subsection{Indene, $c$-C$_9$H$_8$}

The high abundance derived for this aromatic molecule, in fact the first pure
polycyclic aromatic hydrocarbon found in space,
also challenges to find out an efficient formation route in TMC-1. Such a 
route should involve a fast reaction between two abundant species. Potential 
formation reactions like $c$-C$_6$H$_6$ + CH$_2$CCH and $c$-C$_6$H$_5$ + C$_3$H$_4$ have 
activation barriers \citep{Kislov2007,Vereecken2003,Mebel2017}. 
Reactions such as $c$-C$_5$H$_6$ + C$_4$H$_3$ or $c$-C$_5$H$_5$ + C$_4$H$_4$ 
could be efficient producing indene.
\cite{Doddipatla2021} suggest that indene can be formed through the 
barrierless reaction between the radical CH and styrene ($c$-C$_6$H$_5$C$_2$H$_3$).
However, it is unknown whether the precursor styrene is abundant enough in TMC-1, 
and it is not obvious which formation pathway can lead to it. 
\cite{Doddipatla2021} present a tentative synthetic route, from which they calculate 
abundances of up to 10$^{-12}$ for indene, well below the value derived from observations 
here. It is clear that further research is needed to explain the presence of indene in 
TMC-1 with an abundance as high as 1.6\,$\times$10$^{-9}$ relative to H$_2$.

\section{Conclusions}

We report the detection of $c$-C$_3$HCCH, cyclopentadiene, and indene in TMC-1. While the 
observed abundance of $c$-C$_3$HCCH, a few 10$^{-11}$ relative to H$_2$, can be accounted 
for by the reaction between CCH and $c$-C$_3$H$_2$, the fairly high abundances found for 
cyclopentadiene and indene, above 10$^{-9}$, are difficult to explain by currently proposed 
formation routes. The detection of very abundant cyclopentadiene and indene should promote 
further research to elucidate which plausible bottom-up mechanisms of formation of such 
complex aromatic hydrocarbons can be at work in cold dense clouds like TMC-1.

\begin{acknowledgements}

We thank ERC for funding
through grant ERC-2013-Syg-610256-NANOCOSMOS. 
We also thank Ministerio de Ciencia e Innovaci\'on of Spain (MICIU) for funding support through projects
AYA2016-75066-C2-1-P, PID2019-106110GB-I00, PID2019-107115GB-C21 / AEI / 10.13039/501100011033, and
PID2019-106235GB-I00. M.A. thanks MICIU for grant 
RyC-2014-16277. We would like to thank our referee, S. Yamamoto, for his useful comments and
suggestions.

\end{acknowledgements}

\normalsize

\begin{appendix}
\section{Line parameters of $c$-C$_3$HCCH, $c$-C$_5$H$_6$, and $c$-C$_9$H$_8$}
\label{line_parameters}
Line parameters for the different molecules studied in this work were obtained by fitting a Gaussian line
profile to the observed data. A window of $\pm$ 15 \kms\, around the v$_{LSR}$ of the source has been
considered for each transition. The derived line parameters for the three 
molecular species discovered in
this work are given in Table \ref{obs_line_parameters}.

\onecolumn
\begin{longtable}{clcccc}
\caption{Observed line parameters for c-C3HCCH in TMC-1} \label{obs_line_parameters}\\
\hline \hline
$J_{K_a,K_c}$        &$\nu_{obs}$~$^a$    & $\int T_A^* dv$~$^b$ & $\Delta v$~$^c$ & $T_A^*$$^d$&  \\
                   &  (MHz)             & (mK\,km\,s$^{-1}$)   & (km\,s$^{-1}$)  & (mK)   &  \\
\hline
\endfirsthead
\caption{continued.}\\
\hline \hline
$J_{K_a,K_c}$        &$\nu_{obs}$~$^a$    & $\int T_A^* dv$~$^b$ & $\Delta v$~$^c$ & $T_A^*$$^d$&  \\
                   &  (MHz)             & (mK\,km\,s$^{-1}$)   & (km\,s$^{-1}$)  & (mK)   &  \\
            
\hline
\endhead
\hline
\endfoot

$c$-C$_3$HCCH        &                    &                      &        
         &        & \\
$ 1_{1,0}-1_{0,1}$ & 31525.012$\pm$0.03 & 1.30$\pm$0.23& 1.35$\pm$0.28& 0.90$\pm$0.20&  \\ 
$ 2_{1,1}-2_{0,2}$ & 31838.475$\pm$0.03 & 1.98$\pm$0.34& 1.34$\pm$0.20& 1.38$\pm$0.24&  \\ 
$ 5_{1,5}-4_{1,4}$ & 31904.045$\pm$0.03 & 0.53$\pm$0.18& 0.76$\pm$0.35& 0.65$\pm$0.24& A\\ 
$ 3_{1,2}-3_{0,3}$ & 32312.929$\pm$0.03 & 0.87$\pm$0.16& 0.87$\pm$0.17& 0.93$\pm$0.20& A\\ 
$ 9_{0,9}-8_{1,8}$ & 32429.470$\pm$0.03 & 0.88$\pm$0.20& 1.13$\pm$0.28& 0.73$\pm$0.21&  \\ 
$ 5_{0,5}-4_{0,4}$ & 32646.191$\pm$0.03 & 1.34$\pm$0.19& 1.32$\pm$0.24& 0.95$\pm$0.18&  \\ 
$ 4_{1,3}-4_{0,4}$ & 32953.663$\pm$0.03 & 1.07$\pm$0.22& 1.16$\pm$0.33& 0.87$\pm$0.22&  \\ 
$ 5_{1,4}-4_{1,3}$ & 33460.002$\pm$0.03 & 0.70$\pm$0.20& 0.40$\pm$0.23& 1.35$\pm$0.22& B\\ 
$ 5_{1,4}-5_{0,5}$ & 33767.455$\pm$0.03 & 1.24$\pm$0.23& 1.18$\pm$0.24& 0.88$\pm$0.21&  \\ 
$ 6_{1,5}-6_{0,6}$ & 34762.885$\pm$0.03 & 1.09$\pm$0.25& 1.37$\pm$0.38& 0.75$\pm$0.21&  \\ 
$ 7_{1,6}-7_{0,7}$ & 35950.042$\pm$0.02 &              &              &              & C\\
$ 8_{1,7}-8_{0,8}$ & 37340.311$\pm$0.02 &              &              &              & D\\
$ 1_{1,1}-0_{0,0}$ & 37752.281$\pm$0.01 &              &              &              & E\\
$ 6_{1,6}-5_{1,5}$ & 38278.543$\pm$0.01 &              &              &              & C\\
$ 9_{1,8}-9_{0,9}$ & 38946.462$\pm$0.03 &              &              &              & E\\
$ 6_{0,6}-5_{0,5}$ & 39149.938$\pm$0.03 & 0.75$\pm$0.15& 0.87$\pm$0.16& 0.80$\pm$0.21&  \\ 
$10_{0,10}-9_{1,9}$& 40056.165$\pm$0.03 & 1.06$\pm$0.18& 0.81$\pm$0.20& 1.23$\pm$0.21&  \\ 
$ 2_{1,2}-1_{0,1}$ & 43979.464$\pm$0.03 & 0.98$\pm$0.18& 0.96$\pm$0.19& 0.95$\pm$0.26&  \\ 
\hline
$c$-C$_5$H$_6$$^f$      &                   &              &          &              & \\
$3_{1,2}-2_{2,1}$     &37454.017$\pm$0.02 & 1.18$\pm$0.22& 0.79$\pm$0.16& 1.40$\pm$0.30& \\
$3_{2,2}-2_{1,1}$     &38092.525$\pm$0.02 & 0.85$\pm$0.17& 0.68$\pm$0.15& 1.17$\pm$0.24& \\
$4_{0,4}-3_{1,3}$     &38224.588$^e$      &              &              &              & F\\
$4_{1,4}-3_{0,3}$     &38224.787$\pm$0.02 & 0.75$\pm$0.14& 0.68$\pm$0.15& 1.03$\pm$0.20& \\
$4_{1,3}-3_{2,2}$     &46314.423$^e$      &              &              & $\leq$1.5    & \\
$4_{2,3}-3_{1,2}$     &46352.738$\pm$0.02 & 1.37$\pm$0.26& 0.56$\pm$0.12& 2.33$\pm$0.47& \\
$3_{3,1}-2_{2,0}$     &46499.116$\pm$0.02 & 0.91$\pm$0.18& 0.48$\pm$0.09& 1.78$\pm$0.36&\\
$5_{0,5}-4_{1,4}$     &46767.415$\pm$0.02 & 3.49$\pm$0.23& 0.62$\pm$0.05& 5.33$\pm$0.39& G\\
\hline
$c$-C$_9$H$_8$$^f$      &                   &              &              &              & \\
$11_{4, 7}-10_{4, 6}$ &31364.868$\pm$0.02 & 0.47$\pm$0.20& 1.12$\pm$0.39& 0.40$\pm$0.20& \\
$12_{3,10}-11_{3, 9}$ &32083.322$\pm$0.01 & 0.80$\pm$0.18& 1.11$\pm$0.32& 0.68$\pm$0.18& \\
$11_{3, 8}-10_{3, 7}$ &32171.624$\pm$0.01 & 0.83$\pm$0.28& 0.89$\pm$0.33& 0.88$\pm$0.20& \\
$13_{2,12}-12_{2,11}$ &32481.036$\pm$0.01 & 0.74$\pm$0.16& 0.83$\pm$0.18& 0.84$\pm$0.21& \\
$14_{0,14}-13_{1,13}$ &32526.303$\pm$0.01 & 0.94$\pm$0.20& 1.01$\pm$0.24& 0.87$\pm$0.18& \\
$14_{1,14}-13_{1,13}$ &32527.242$\pm$0.01 & 0.78$\pm$0.21& 1.05$\pm$0.33& 0.70$\pm$0.18& \\
$14_{0,14}-13_{0,13}$ &32528.322$\pm$0.01 & 1.14$\pm$0.21& 1.65$\pm$0.15& 0.65$\pm$0.18& \\
$14_{1,14}-13_{0,13}$ &32529.269$\pm$0.01 & 0.61$\pm$0.19& 0.60$\pm$0.24& 0.97$\pm$0.20& \\
$13_{1,12}-12_{1,11}$ &32555.319$\pm$0.03 & 1.15$\pm$0.26& 1.87$\pm$0.45& 0.57$\pm$0.20&A\\
$12_{2,10}-11_{2, 9}$ &33087.091$\pm$0.02 & 0.21$\pm$0.11& 0.48$\pm$0.20& 0.40$\pm$0.16& \\
$12_{6, 7}-11_{6, 6}$ &33104.538$\pm$0.01 &              &              & $\le$0.48    & \\
$12_{4, 9}-11_{4, 8}$ &33105.876$\pm$0.02 & 0.30$\pm$0.16& 0.40$\pm$0.20& 0.45$\pm$0.16& \\
$12_{6, 6}-11_{6, 5}$ &33124.468$\pm$0.02 & 0.56$\pm$0.17& 0.84$\pm$0.19& 0.70$\pm$0.19&A\\
$12_{5, 8}-11_{5, 7}$ &33262.106$\pm$0.01 & 0.91$\pm$0.18& 0.81$\pm$0.20& 1.05$\pm$0.22& \\
$12_{5, 7}-11_{5, 6}$ &33524.184$\pm$0.02 & 0.26$\pm$0.10& 0.34$\pm$0.20& 0.72$\pm$0.18& \\
$13_{3,11}-12_{3,10}$ &34484.248$\pm$0.01 & 0.72$\pm$0.16& 0.96$\pm$0.26& 0.70$\pm$0.18& \\
$12_{4, 8}-11_{4, 7}$ &34608.352$\pm$0.01 & 0.88$\pm$0.16& 1.11$\pm$0.22& 0.75$\pm$0.18& \\
$14_{2,13}-13_{2,12}$ &34736.637$\pm$0.01 & 0.60$\pm$0.19& 1.02$\pm$0.36& 0.55$\pm$0.21& \\
$15_{0,15}-14_{1,14}$ &34771.228$\pm$0.02 &              &              & $\le$ 0.45   & \\
$15_{1,15}-14_{1,14}$ &34771.684$\pm$0.02 & 0.30$\pm$0.08& 0.54$\pm$0.17& 0.53$\pm$0.15& \\
$15_{0,15}-14_{0,14}$ &34772.190$\pm$0.02 & 0.49$\pm$0.32& 0.94$\pm$0.33& 0.49$\pm$0.15&A\\
$15_{1,15}-14_{0,14}$ &34772.612$\pm$0.02 &              &              &  $\le$0.45   & \\
$14_{1,13}-13_{1,12}$ &34776.320$\pm$0.02 & 0.40$\pm$0.08& 0.33$\pm$0.15& 1.13$\pm$0.17& \\
$12_{3, 9}-11_{3, 8}$ &35011.580$\pm$0.01 & 0.57$\pm$0.13& 0.63$\pm$0.15& 0.85$\pm$0.19& \\
$13_{2,11}-12_{2,10}$ &35211.250$\pm$0.01 & 0.43$\pm$0.08& 0.49$\pm$0.22& 0.82$\pm$0.17& \\
$13_{4,10}-12_{4, 9}$ &35766.651$\pm$0.02 & 0.68$\pm$0.17& 1.12$\pm$0.26& 0.57$\pm$0.19& \\
$13_{6, 8}-12_{6, 7}$ &35959.203$\pm$0.01 &              &              & $\le$ 0.60   & \\
$13_{6, 7}-12_{6, 6}$ &36008.349$\pm$0.02 & 0.64$\pm$0.15& 1.00$\pm$0.34& 0.60$\pm$0.20&A\\ 
$13_{5, 9}-12_{5, 8}$ &36101.117$\pm$0.02 & 0.64$\pm$0.17& 1.10$\pm$0.40& 0.47$\pm$0.16& \\
$13_{5, 8}-12_{5, 7}$ &36605.949$\pm$0.01 & 0.71$\pm$0.15& 0.71$\pm$0.15& 0.94$\pm$0.21& \\
$14_{3,12}-13_{3,11}$ &36835.520$\pm$0.01 & 0.80$\pm$0.16& 1.02$\pm$0.25& 0.74$\pm$0.18& \\
$15_{2,14}-14_{2,13}$ &36986.145$\pm$0.02 & 0.32$\pm$0.11& 0.45$\pm$0.16& 0.59$\pm$0.19& \\
$15_{1,14}-14_{1,13}$ &37006.756$\pm$0.01 & 0.60$\pm$0.14& 0.65$\pm$0.16& 0.87$\pm$0.19& \\
$16_{1,16}-15_{1,15}$ &37016.001$\pm$0.02 & 0.39$\pm$0.15& 0.77$\pm$0.27& 0.47$\pm$0.20& \\
$16_{0,16}-15_{0,15}$ &37016.269$\pm$0.02 & 0.68$\pm$0.21& 1.05$\pm$0.35& 0.57$\pm$0.20&A\\
$14_{2,12}-13_{2,11}$ &37318.440$\pm$0.01 & 0.30$\pm$0.12& 0.60$\pm$0.20& 0.47$\pm$0.20& \\
$13_{3,10}-12_{3, 9}$ &37665.778$\pm$0.01 & 0.84$\pm$0.16& 0.78$\pm$0.17& 1.01$\pm$0.21& \\
$13_{4, 9}-12_{4, 8}$ &37842.548$\pm$0.01 & 0.41$\pm$0.16& 0.64$\pm$0.21& 0.70$\pm$0.19& \\
$14_{4,11}-13_{4,10}$ &38361.360$\pm$0.01 &              &              &              &C\\
$14_{6, 9}-13_{6, 8}$ &38829.224$\pm$0.01 & 0.87$\pm$0.17& 0.83$\pm$0.16& 1.00$\pm$0.23& \\
$14_{5,10}-13_{5, 9}$ &38918.551$\pm$0.01 &              &              & $\le$0.69    & \\
$14_{6, 8}-13_{6, 7}$ &38939.641$\pm$0.02 &              &              & $\le$0.69    & \\
$15_{3,13}-14_{3,12}$ &39147.950$\pm$0.02 & 0.38$\pm$0.13& 0.48$\pm$0.20& 0.75$\pm$0.23& \\
$16_{2,15}-15_{2,14}$ &39232.293$\pm$0.02 & 1.02$\pm$0.27& 0.68$\pm$0.19& 1.46$\pm$0.24& \\
$16_{1,15}-15_{1,14}$ &39242.748$\pm$0.02 &              &              & $\le$1.6     & \\
$17_{1,17}-16_{1,16}$ &39260.308$\pm$0.02 & 0.71$\pm$0.23& 0.54$\pm$0.18& 1.21$\pm$0.40& \\
$17_{0,17}-16_{0,16}$ &39260.513$\pm$0.02 & 1.04$\pm$0.26& 0.61$\pm$0.15& 1.61$\pm$0.40& \\
$15_{2,13}-14_{2,12}$ &39447.608$\pm$0.03 & 0.54$\pm$0.18& 0.82$\pm$0.29& 0.62$\pm$0.22& \\
$14_{5, 9}-13_{5, 8}$ &39799.463$\pm$0.03 & 0.51$\pm$0.18& 1.10$\pm$0.33& 0.43$\pm$0.20& \\
$14_{3,11}-13_{3,10}$ &40118.461$\pm$0.02 &              &              & $\le$0.65    & \\
$15_{4,12}-14_{4,11}$ &40886.724$\pm$0.03 & 0.58$\pm$0.20& 0.95$\pm$0.35& 0.57$\pm$0.26& \\
$14_{4,10}-13_{4, 9}$ &40983.754$\pm$0.03 & 0.37$\pm$0.17& 0.60$\pm$0.29& 0.58$\pm$0.24& \\
$16_{3,14}-15_{3,13}$ &41432.310$\pm$0.03 &              &              & $\le$0.90    & \\
$17_{2,16}-16_{2,15}$ &41476.748$\pm$0.03 &              &              &              &B\\
$17_{1,16}-16_{1,15}$ &41482.002$\pm$0.03 &              &              & $\le$0.90    & \\
$18_{1,18}-17_{1,17}$ &41504.704$\pm$0.01 & 0.95$\pm$0.20& 0.74$\pm$0.18& 1.21$\pm$0.29&H\\
$18_{0,18}-17_{0,17}$ &41504.704$\pm$0.01 &              &              &              &H\\
$16_{2,14}-15_{2,13}$ &41608.793$\pm$0.02 & 0.63$\pm$0.27& 0.49$\pm$0.17& 1.20$\pm$0.33& \\
$15_{5,11}-14_{5,10}$ &41698.205$\pm$0.02 &              &              & $\le$0.90    & \\
$19_{1,19}-18_{1,18}$ &43748.996$\pm$0.02 & 1.00$\pm$0.30& 0.92$\pm$0.38& 1.09$\pm$0.35&H\\
$19_{0,19}-18_{0,18}$ &43748.996$\pm$0.02 &              &              &              &H\\ 
$20_{1,20}-19_{1,19}$ &45993.359$\pm$0.02 & 1.24$\pm$0.31& 0.64$\pm$0.23& 1.83$\pm$0.42&H\\
$20_{0,20}-19_{0,19}$ &45993.369$\pm$0.02 &              &              &              &H\\
\hline
\end{longtable}
\tablefoot{\\
\tablefoottext{a}{Observed frequency assuming a v$_{LSR}$ of 5.83 \kms.}\\
\tablefoottext{b}{Integrated line intensity in mK\,km\,s$^{-1}$.}\\
\tablefoottext{c}{Linewidth at half intensity derived by fitting a Gaussian function to
the observed line profile (in km\,s$^{-1}$).}\\
\tablefoottext{d}{Antenna temperature in milli Kelvin.}\\
\tablefoottext{e}{Predicted frequency.}\\
\tablefoottext{f}{Measured frequencies assume a v$_{LSR}$ of 5.83 \kms. The difference between observed
 and predicted frequencies are $\pm$15 kHz.}\\
\tablefoottext{A}{Partially blended with another feature. Fit still possible providing reasonable results.}\\
\tablefoottext{B}{Blended with another feature. Fit uncertain.}\\
\tablefoottext{C}{Fully blended with a negative feature produced in the frequency switching folding. Fit unreliable.
  Frequency corresponds to the predicted value from the rotational constants of Table 1.}\\
\tablefoottext{D}{Fully blended with a line from HCCCH$_2$CN. Frequency corresponds to the predicted value
  from the rotational constants of Table \ref{rotational_constants}.}\\
\tablefoottext{E}{A narrow line is clearly visible but it is affected by a negative feature.
  Frequency corresponds to the predicted value from the rotational constants of Table 1.}\\
\tablefoottext{F}{Fully blended with a line from $c$-C$_3$HD. Frequency corresponds to the predicted value.}\\
\tablefoottext{G}{The $5_{1,5}-4_{0,4}$ line, which belongs to the para species, is fully blended
 with the $5_{0,5}-4_{1,4}$ transition of the ortho species. Both lines have similar line strength.}\\
\tablefoottext{H}{Unresolved doublet.}\\
}
\twocolumn

\section{The isomers of $c$-C$_3$HCCH}
\label{Ap_isomers_C3HCCH}
The isomeric family with formula H$_2$C$_5$ is composed by five molecular species.
The nonpolar pentadiynylidene (HCCCCH) and ethynyl cyclopropenylidene ($c$-C$_3$HCCH) are
the most stable isomers, whose energy separation is very small $\thicksim$0.043 eV.
H$_2$CCCCC is a high energy isomer and lies 0.564 eV above the ground. Other two species
are part of this isomeric family, HCC(CH)CC and $c$-C$_3$H$_2$CC, placed at 0.737 and 
0.910 eV respectively, 
above the most stable forms, respectively \citep{Seburg1997}.

The rotational transitions in the 31-50 GHz of HCC(CH)CC, one of the C$_5$H$_2$ isomers, can be predicted with high
accuracy from the laboratory data of \citet{Gottlieb1998}. We obtain a 3$\sigma$ upper limit to its
column density of 2$\times$10$^{10}$ cm$^{-2}$. Note that the dipole moment of this
species is 4.52 D \citep{Gottlieb1998}. An even larger dipole moment has been calculated by the
same authors for the isomer $c$-C$_3$H$_2$CC (8.2 D). For this species we 
derive a 3$\sigma$ upper limit 
to its column density of 9$\times$10$^{9}$ cm$^{-2}$. Finally, the cumulenic species H$_2$C$_5$ has been detected
recently from our data and the results will be published elsewhere 
(Cabezas et al. 2021, in preparation).

An additional derivate of $c$-C$_3$H$_2$ is cyano propenylidene, $c$-C$_3$HCN, which has been
observed in the laboratory by \citet{McCarthy1999}. We searched for it in 
our data and
derive a 3$\sigma$ upper limit to its column density of 2.5$\times$\once.

\section{Improved rotational constants for $c$-C$_3$HCCH and c-C$_9$H$_8$}
\label{improved_rotational_constants}
The frequencies we have measured in TMC-1 can be used to improve the rotational and distortion constants
of $c$-C$_3$HCCH and indene. We have used the fitting code FITWAT described in \citet{Cernicharo2018}. 

Table \ref{rotational_constants}
provides the results obtained by fitting the laboratory data of \citet{Travers1997} 
for c-C$_3$HCCH alone, and those
obtained from a fit to the merged laboratory plus the TMC-1 frequencies. 
An improvement of the uncertainty in
the rotational and distortion constants is obtained. The fit to the laboratory data alone results in exactly
the same constants than those obtained by \citet{Travers1997}. 

Table \ref{rotational_constants_indene} provide the same information but for indene. 
The 50 lines measured in TMC-1 (three of them are unresolved doublets)
provide a significant improvement of the rotational and
distortion constants for this species.

The merged fits are recommended to predict the
frequency of the rotational transitions of both species with uncertainties
between 10 and 200 kHz up to 115 GHz.

\begin{table}
\tiny
\caption{Rotational and distortion constants of $c$-C$_3$HCCH}
\label{rotational_constants}
\centering
\begin{tabular}{lcc}
\hline
Constant            & Laboratory$^a$      & This work         \\
\hline
$A$ (MHz)           &  34638.7012(27)   &34638.7023(26)   \\
$B$ (MHz)           &   3424.87685(43)  & 3424.87678(42)  \\
$C$ (MHz)           &   3113.63856(50)  & 3113.63900(44)  \\
$\Delta_J$ (kHz)    &      0.2877(80)   &   0.2918(74)    \\
$\Delta_{JK}$ (kHz) &     29.53(24)     &  29.72(21)      \\
\hline
Number of lines     & 13                  & 26                \\
$\sigma$(kHz)       & 2.1                 & 26                \\
$J_{max}$, $K_{max}$& 7, 1               & 10,1             \\
$\nu_{max}$ (GHz)   & 26.130              & 43.979            \\
\hline
\hline
\end{tabular}
\tablefoot{\\
Values between parenthesis correspond to the uncertainties of the parameters 
in units of the least significant digits.\\
\tablefoottext{a}{Laboratory frequencies from \citet{Travers1997}.}\\
\\
}
\end{table}

\begin{table}
\tiny
\caption{Rotational and distortion constants of $c$-C$_9$H$_8$}
\label{rotational_constants_indene}
\centering
\begin{tabular}{lcc}
\hline
Constant            & Laboratory$^a$     & This work       \\
\hline
$A$ (MHz)           &   3775.048(15)     & 3775.0469(80)   \\
$B$ (MHz)           &   1580.8656(21)    & 1580.86511(79)  \\
$C$ (MHz)           &   1122.2460(18)    & 1122.24773(51)  \\
$\Delta_J$ (kHz)    &      0.0326(38)    &   0.03349(97)    \\
$\Delta_{JK}$ (kHz) &      0.050(11)     &   0.0517(73)    \\
$\Delta_K$ (kHz)    &      0.37(12)      &   0.329(82)     \\
$\delta_J$ (kHz)    &      0.0117(12)    &   0.01068(51)   \\
$\delta_K$  (kHz)   &      0.068(19)     &   0.0775(94)    \\
\hline
Number of lines     & 78                 & 128                \\
$\sigma$(kHz)       & 67                 & 54                \\
$J_{max}$, $K_{max}$& 30,13              & 30,13             \\
$\nu_{max}$ (GHz)   & 39.3               & 45.99             \\
\hline
\hline
\end{tabular}
\tablefoot{\\
Values between parenthesis correspond to the uncertainties of the parameters 
in units of the least significant digits.\\
\tablefoottext{a}{Laboratory frequencies from \citet{Li1979,Caminati1993}.}\\
\\
}
\end{table}

\section{Column density of $c$-C$_3$H$_2$}
\label{Ap_C3H2}
\begin{figure}[]
\centering
\includegraphics[scale=0.65,angle=0]{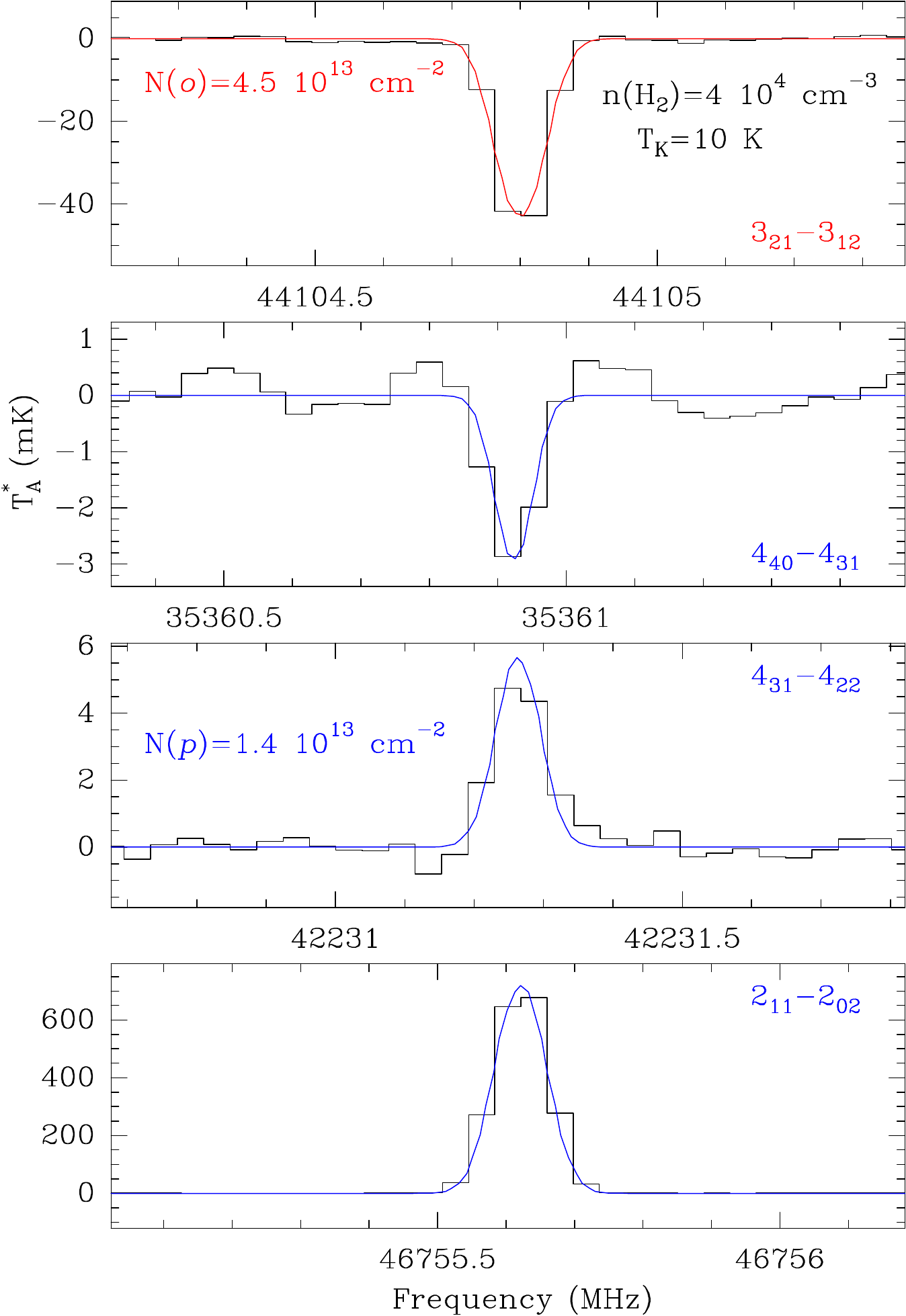}
\caption{Observed transitions of $c$-C$_3$H$_2$ in TMC-1.
The abscissa corresponds to the rest frequency of the lines assuming a
local standard of rest velocity of the source of 5.77 km s$^{-1}$. 
The ordinate is the antenna temperature, corrected for atmospheric and telescope losses, in mK.
The quantum numbers for each transition are indicated
in the upper or lower right corner of the corresponding panel.
The red line shows the computed synthetic spectrum for this species (see text). 
Blue labels indicate the multiplicative factor applied, when needed, to the model to
match the intensity of the observed lines.
}
\label{fig_c3h2}
\end{figure}

The polar carbene ring molecule $c$-C$_3$H$_2$ is widespread in interstellar and circumstellar clouds
\citep{Matthews1985}. For the physical conditions of TMC-1 we expect only 
a few lines of this molecule to
be strong enough within the 
frequency coverage of our survey. In fact, only one ortho line, the $3_{21}-3_{12}$ at 44104.777 MHz
has an upper energy level below 20 K. This line appears in our data in absorption as shown
in Fig. \ref{fig_c3h2}. We have checked that the absorption is real and not produced by
a negative feature created during the folding of the frequency switching data by analyzing separately 
the two frequency throws used during the observations. This line was predicted to be in absorption by \citet{Avery1989}
for the typical densities of cold dark clouds, but to the best of our knowledge it has not been reported previously.
The other two lines of the ortho species in our survey have upper level energies around 45 K and are not detected.
Absorption from $c$-C$_3$H$_2$ in cold dark clouds has been previously reported for
its para line $2_{20}-2_{11}$ at 21.6 GHz \citep{Matthews1986}. For the para species three lines are detected in
our data as shown in Fig. \ref{fig_c3h2}. They are the $4_{40}-4_{31}$ at 
35.36 GHz, the $4_{31}-4_{22}$ at 42.231 GHz, 
and the $2_{11}-2_{02}$ at 46.755 GHz.
The first one appears also in absorption. We have checked, as for $3_{21}-3_{12}$ ortho transition, 
that the absortion is real. The other two para lines are in emission.

Collisional rates between $c$-C$_3$H$_2$ and He adapted to the low temperatures of cold dark clouds
are available from \cite{Avery1989} and \cite{Khalifa2019}. 
We performed Large Velocity Gradient (LVG) calculations for ortho and para cyclopropenylidene species
by varying the density and the column density and by adopting the most recent set of collisional rates.
The radius of the source is fixed to 40$''$ \citep{Fosse2001} and the linewidth to 0.5 \kms.
The best fit to the three observed para lines is obtained for $n$(H$_2$)\,=(\,4.0$\pm$0.5)$\times$10$^4$
cm$^{-2}$, and $N$($p$-C$_3$H$_2$)=(1.4\,$\pm$\,0.3)\,$\times$\,\trece. 
Adopting the derived density for the para species, then 
the ortho absorption line can be reproduced for a column density $N$($o$-C$_3$H$_2$)\,=\,4.5\,$\times$\,\trece. 
Hence, the ortho to para abundance ratio of $c$-C$_3$H$_2$ is $\sim$\,3, which is the expected value from
the spin degeneracy of the two species. The total column density of 
cyclopropenylidene is 5.9\,$\times$\,\trece, in very good
agreement with the value derived by \citet{Fosse2001} of 5.8\,$\times$\,\trece. Nevertheless, the derived abundances
rely on the accuracy of the collisional rate coefficients. The synthetic spectrum computed from these
parameters is shown by the red (ortho) and blue (para) lines in Fig. \ref{fig_c3h2}, 
and reproduces nicely the ortho and para absorption lines, and the
two para emission lines. If we use the collisional rates calculated by \citet{Avery1989}, a reasonable fit can be
obtained reproducing the two absorption lines. In this case the derived volume density is $n$(H$_2$)\,=\,(2.8\,$\pm$\,0.5)\,$\times$\,10$^4$
cm$^{-3}$ and the best fit to the intensities correspond to $N$($o$-C$_3$H$_2$)\,=\,(2.8\,$\pm$\,0.5)\,$\times$\,\trece\, and
$N$($p$-C$_3$H$_2$)\,=\,(3.2\,$\pm$\,0.5)\,$\times$\,\trece. With these 
less accurate collisional rates the ortho/para ratio is $\sim$1.
Moreover, deviations of up to 20\% between calculated and observed intensities are observed for the best model fit.
These deviations are not observed when we use the most recent calculations of \cite{Khalifa2019}, which are based
in a more precise determination of the potential energy surface of the $c$-C$_3$H$_2$/He system.
For both set of rates the derived total column density of $c$-C$_3$H$_2$ is practically identical, $\sim$\,6\,$\times$\,\trece.
In view of the complex system of absorption and emission lines shown by this molecular species,
collisional rates using H$_2$ as collider are highly desirable.

\section{Column density of $c$-C$_3$H}
\label{Ap_C3H}
\begin{figure*}[]
\centering
\includegraphics[scale=0.85,angle=0]{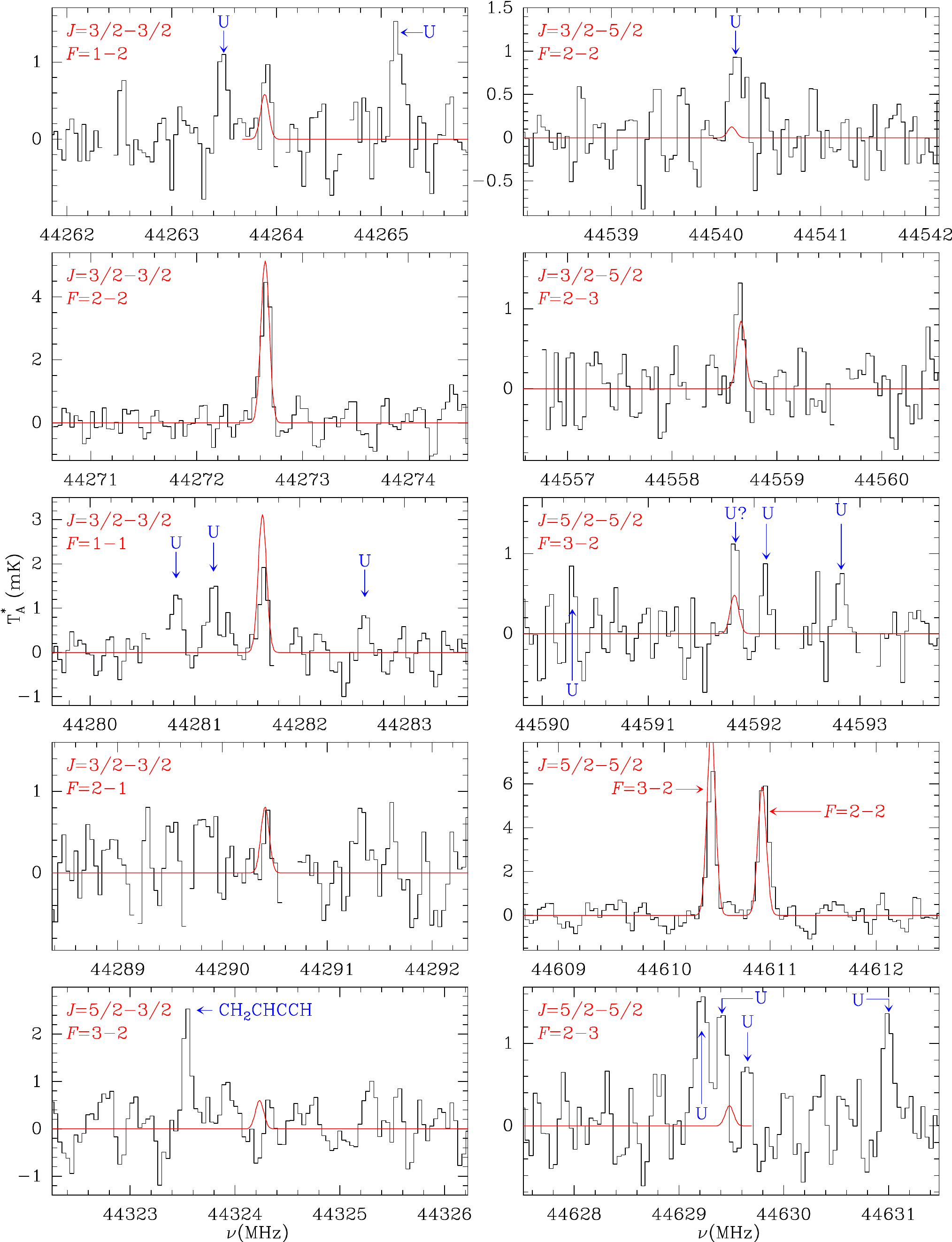}
\caption{Fine and hyperfine components of the $2_{1,1}-1_{1,1}$ transition of $c$-C$_3$H in TMC-1 as
observed with the Yebes 40m radiotelescope.
The abscissa corresponds to the rest frequency of the lines assuming a
local standard of rest velocity of the source of 5.83 km s$^{-1}$. 
The ordinate is the antenna temperature, corrected for atmospheric and telescope losses, in mK.
The quantum numbers for each transition are indicated
in the upper or lower right corner of the corresponding panel.
The red line shows the computed synthetic spectrum for this species 
for a column density of 6.2\,$\times$\,10$^{12}$ cm$^{-2}$ and $T_r$\,=\,3.7\,K (see text). 
}
\label{fig_c-c3h}

\end{figure*}

Cyclopropanediylidenyl, $c$-C$_3$H, has been detected towards several astrophysical
environments including cold dark clouds, evolved stars, and translucent molecular clouds
\citep{Yamamoto1987,Mangum1990,Turner2000,Cernicharo2000}. Several lines pertinining to the
fine and hyperfine structure of the $2_{1,1}-2_{1,2}$ rotational transitions of this species are
within our line survey of TMC-1. The data for all the strongest components, and some of the weak ones, are
shown in Fig. \ref{fig_c-c3h}. Several of the panels of this Figure show a significant number of unknown
features. This is an important drawback for any stacking procedure used to detect molecules producing
intensities below 1\,mK for their rotational transitions. At this level of sensitivity TMC-1 cannot
be considered as a line-poor source and detections have to be performed using the standard procedure
of line-by-line detection.

\begin{figure*}[]
\centering
\includegraphics[scale=0.85,angle=0]{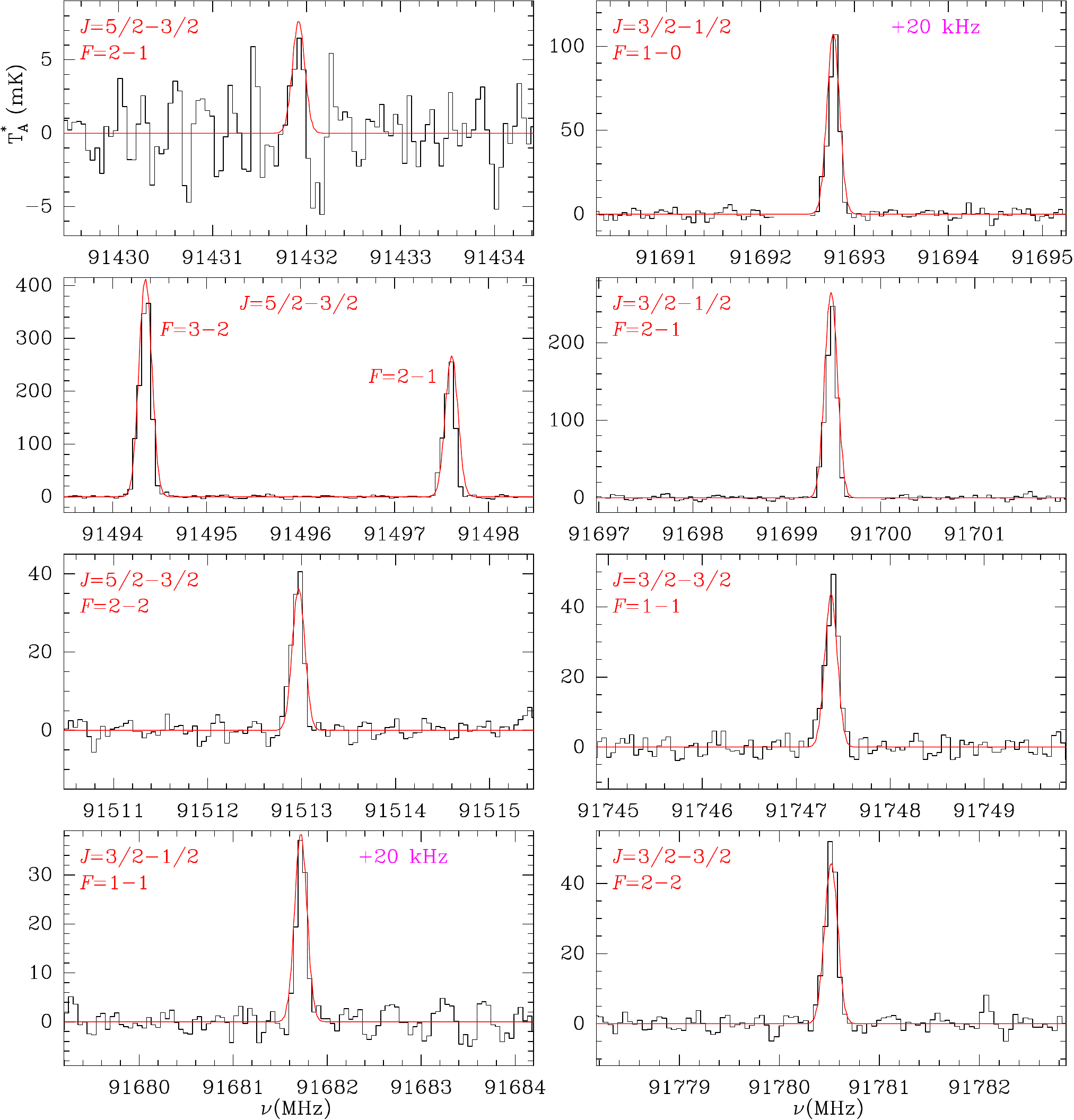}
\caption{Fine and hyperfine components of the $2_{1,2}-1_{1,1}$ transition of $c$-C$_3$H in TMC-1 as
observed with the IRAM 30m radiotelescope (data from \citealt{Marcelino2007}).
The abscissa corresponds to the rest frequency of the lines assuming a
local standard of rest velocity of the source of 5.83 km s$^{-1}$. 
The ordinate is the antenna temperature, corrected for atmospheric and telescope losses, in mK.
The quantum numbers for each transition are indicated
in the upper or lower right corner of the corresponding panel.
The red line shows the computed synthetic spectrum for this species 
for a column density of 6.2\,$\times$\,10$^{12}$ cm$^{-2}$ and $T_r$\,=\,5\,K (see text). 
Violet labels indicate the observed minus calculated frequencies when
larger than 5 kHz.
}
\label{fig_c-c3h_3mm}
\end{figure*}

With only one transition we have to assume a rotational temperature in
order to derive a column density for the molecule. 
The energy of the upper level of the different fine and hyperfine components of the $2_{1,1}-2_{1,2}$ transition
is $\sim$\,6.5 K. Hence,
the column density will depend slightly on the adopted value of $T_r$ (see, e.g., the
error analysis carried out by \citealt{Cernicharo2021c} for HC$_3$S$^+$). 
Assuming $T_r$\,=\,10\,K, a source diameter
of 80$''$ and a linewidth of 0.6 \kms,
we obtain a column density for $c$-C$_3$H of (1.2\,$\pm$\,0.2)\,$\times$\,\doce. This value is a factor five smaller
than the one derived by \cite{Yamamoto1987}, who observed the $2_{1,2}-1_{1,1}$ transition at $\sim$\,91.5 GHz
with the Nobeyama radiotelescope. They assumed a rotational temperature of 5 K. 
In order to understand this large difference we have used
the data gathered with the IRAM 30m telescope
during the 3 mm line survey of TMC-1 \citep{Marcelino2007}. We obtain intensities for the $2_{1,2}-1_{1,1}$ hyperfine 
components that are similar to those of \citet{Yamamoto1987}. The 3 mm data are shown in Fig. \ref{fig_c-c3h_3mm}.
Assuming an identical rotational temperature for the two transitions it is not possible to
get a reliable fit to the observed intensities. In fact, the 3 mm lines are well explained
with a column density of (6.2\,$\pm$\,0.4)\,$\times$\,10$^{12}$ cm$^{-2}$ 
for $T_r$\,=\,5\,K, a very similar value 
to that of \cite{Yamamoto1987}. However, the rotational temperature has to be decreased to 3.7 K to
explain with the same column density the observed intensities for the $2_{1,1}-2_{1,2}$. The two transitions
share the same lower level (the fundamental one), but they could have significant different excitation
temperatures if the collisional rates from the ground to the two upper levels are different. Unfortunately
no collisional rates for $c$-C$_3$H are available in the literature.


\end{appendix}

\end{document}